\documentclass[journal,review,submit,moreauthors,12pt,a4paper]{mdpi} 
\usepackage{amsmath,amssymb}
\firstpage{1} 
\articlenumber{x}
\doinum{ --}
\pubvolume{xx}
\pubyear{2018}
\copyrightyear{2018}
\history{Received: date; Accepted: date; Published: date}
\usepackage[T1]{fontenc}
\usepackage{lmodern}
\def\apj{\rm ApJ}
\def\apjl{\rm ApJL}
\def\apjs{\rm ApJS}
\def\aj{\rm AJ}
\def\mnras{\rm MNRAS}

\def\nat{\rm Nature}

\def\pasp{\rm PASP}
\def\aap{\rm A\&A}

\def\aapr{\rm AAPR}
\def\pasa{\rm PASA}
\def\ssr{\rm SSR}
\def\apss{\rm APSS}
\def\amp{\rm AMP}

\Title{Radio galaxies - the TeV challenge}

\Author{Bindu Rani $^{1,\dagger}$\orcidA{}}

\AuthorNames{Bindu Rani}

\address{%
$^{1}$ \quad NASA Goddard Space Flight Center, Greenbelt, MD, 20771, USA}

\corres{email: bindu.rani@nasa.gov}

\firstnote{NPP Fellow} 

\abstract{ 
Over the past decade, our knowledge of the $\gamma$-ray sky has been revolutionized by 
ground- and space-based observatories  by detecting  photons up to several 
hundreds of tera-electron volt (TeV) energies. A major population 
of the $\gamma$-ray bright objects are active galactic nuclei  (AGN) with their relativistic 
jets pointed along our line-of-sight. Gamma-ray emission  is also detected 
from  nearby mis-aligned AGN such as radio galaxies. While the TeV-detected radio galaxies ({\bf $TeVRad$}) only 
form a small fraction of the $\gamma$-ray detected AGN, their multi-wavelength study 
offers a unique opportunity to probe and pinpoint the high-energy emission processes and 
sites. Even in the absence of substantial Doppler beaming $TeVRad$ are extremely bright 
objects in the TeV sky (luminosities detected up to $10^{45}~erg~s^{-1}$), and exhibit 
flux variations on timescales shorter than the 
event-horizon scales (flux doubling timescale less than 5 minutes).  Thanks to the recent  
advancement in the imaging capabilities  of 
 high-resolution radio interferometry (millimeter very long baseline interferometry, mm-VLBI), one can probe the scales 
down to less than 10 gravitational radii in $TeVRad$, making it possible not only to 
test  jet launching models but also to pinpoint the high-energy emission sites and 
to unravel the emission mechanisms. 
This review provides an overview of the high-energy observations 
of  $TeVRad$ with a focus on the emitting sites and radiation processes. 
Some recent approaches in simulations are also  
sketched. Observations by the  near-future facilities like Cherenkov Telescope Array, 
short millimeter-VLBI, and high-energy polarimetry instruments  will be crucial for 
discriminating the competing high-energy emission models. 
}  

\begin{document}

\section{Introduction}
An exciting discovery enabled by space- and ground-based high-energy observations is the detection of 
$\gamma$-rays from over 3,000 extragalactic sources. Active galactic nuclei (AGN, powered by 
accretion onto a few million to several billion solar mass black holes) form a major population of the 
$\gamma$-ray sky. 
AGN often produce collimated outflows, called relativistic jets (see Fig. 1), whose beams of radiation 
pierce through the Universe and reach us from the past, offering a unique 
opportunity to study the Universe when it was an order of magnitude younger than today. 
Despite the intensive study of AGN, 
the location and origin of their  $\gamma$-rays remains a mystery.
Some studies imply the 
emission region is located  close to the central black hole \citep{rani2013, rani2014, algaba2018, 3c84_tev, acciari2009}; while there are 
counterexamples which argue for $\gamma$-rays coming from  farther out in the 
jet \citep{agudo2011, jorstad2013, hodgson2018}.
Extremely bright $\gamma$-ray flashes in AGN  on time scales of minutes to hours 
have attracted the attention of the entire astronomical community, as this suggests 
that the particles that produce  $\gamma$-rays can be accelerated with enormous efficiency in tiny  
regions within more extended  sources.  
The question of what powers $\gamma$-ray AGN flares is ultimately related to the  energy 
dissipation mechanism at work in their relativistic jets.

\smallskip

AGN jets are among 
the largest and most efficient 
particle accelerators in the Universe. However, where and how these particles produce 
high-energy radiation is still an open question. What are the particle acceleration 
processes, which type of particles are the primary energy drivers, and what are 
the high-energy interaction processes are  key puzzles in high-energy astrophysics. 

\smallskip

\begin{figure}
\centering
\includegraphics[scale=0.4, trim = 1 200 1 100, clip]{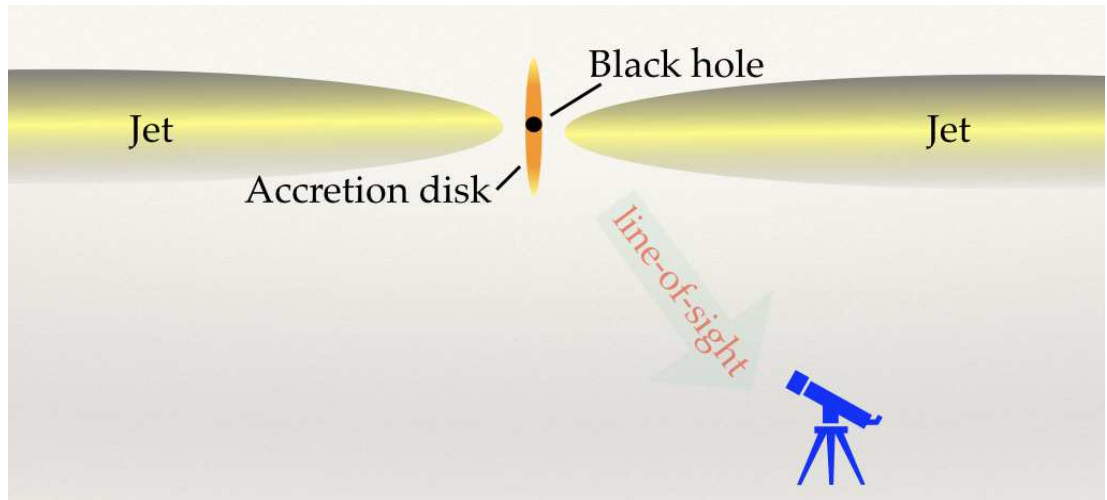}
\caption{Sketch representing the mis-aligned radio emitting bipolar jets of a radio galaxy (not to scale). 
Jets typically extend up to a few hundreds of kilo-parsec to mega-parsec scales. }
\label{fig1}
\end{figure}

AGN having their jets pointed along our line-of-sight, called blazars, are   the  
majority of the sources detected by  the high-energy ground- and space-based telescopes. 
This goes along with the substantial Doppler boosting of their intrinsic emission. The radiation is  beamed 
in our direction making these sources appear to be extremely bright and detectable even at larger redshifts. 
Blazars come in two flavors:  BL Lacertae objects (BL~Lacs) and flat spectrum 
radio quasars (FSRQs). 
In spite of the dissimilarity of their optical spectra -- FSRQs show strong broad emission 
lines, while BL~Lacs have only weak or no emission lines in their optical spectra -- 
they share the same peculiar continuum properties (strong variability and high polarization).   
Most of the  radiation we observe from these blazars is dominated 
by emission from the jet; however characteristic signatures of emission from the 
immediate vicinity of the central black hole (accretion disk or corona) have been also observed 
especially in FSRQs.

The radio to optical emission (and X-rays in some cases) from blazars  is mostly synchrotron radiation from the jet, produced by 
relativistic electrons interacting with the jet's magnetic field. High-energy 
emission, which often dominates the
spectral energy distribution (SED), is from inverse-Compton scattering of 
thermal photons (from the accretion disk, the broad-line region, or the molecular torus) and/or non-thermal photons
(synchrotron emission from the jet)  by the jet's relativistic electrons in case of leptonic models
\citep{rani2013, bottcher2013, bottcher2007, rybicki1986, dermer2009}. 
Gamma-ray emission could also be  produced via photo-hadron interactions or by proton or charged  $\mu$/$\pi$ synchrotron emission 
and subsequent cascades in the framework of hadronic models \citep{bottcher2013, zhang2013, dermer2009}.
Detailed investigations of multi-wavelength flux and spectral variability of individual 
sources including cross-band (radio, optical, X-ray, $\gamma$-ray, and polarization), 
relative timing analysis of outbursts, and/or very long baseline interferometry 
(VLBI) component ejection/kinematics 
suggest that high-energy emission especially {\it $\gamma$ rays are associated with the 
compact regions of relativistic jets energized by the central SMBHs} 
\citep[e.g.,][]{agudo2011, jorstad2001, jorstad2010, marscher2010, marscher2011, 
schinzel2012, rani2013, rani2014, rani2015, orienti2013, raiteri2013}.
However, many critical details of particle acceleration 
mechanisms and radiation processes are still unknown.

Mis-aligned or non-blazar AGN such as radio galaxies or Narrow line Seyfert 1 galaxies, 
although representing a 
small fraction ($<$2$\%$) of the $\gamma$-ray detected sources, have emerged as an interesting 
class of $\gamma$-ray emitting 
AGN.   
Radio galaxies are a particularly interesting class of objects to understand several 
aspects of AGN physics and high-energy emission, i.e.\ how the relativistic 
outflows are launched and driven, what drives the particle acceleration, how and 
where high-energy emission is produced in jets. Several of these key 
questions can be addressed  via studying the TeV-emitting radio galaxies ($TeVRad$ hereafter). 
This review paper summarizes the key observations of $TeVRad$, with an 
overview about the possible particle acceleration processes and emission mechanisms, and 
concentrates on the jet structure and multi-wavelength variability rather than spectral modeling, 
which is a separate topic.  The interested reader is referred to the review article by \citet{rieger2018}.

\begin{table}
\caption{Radio galaxies detected at TeV energies}
\begin{tabular}{llccc}
\toprule
\textbf{Source}	& Type & \textbf{Redshift (Distance in Mpc)}	& \textbf{M$_{BH}$ (M$_{\odot}$}) 
 & \textbf{L$_{VHE}$ (erg s$^{-1}$)}\\
\midrule
Centaurus A &FR1	& 0.00183  (3.7) \citep{harris2010}       & 5$\times10^{7}$ \citep{bettoni2003}    &  $10^{40}$     \\
M87	       &FR1 & 0.0044    (16) \citep{bird2010}       & 6$\times10^{9}$  \citep{walsh2013}   &  $10^{41}$     \\
3C 84	   &FR1 & 0.0177	(71) \citep{strauss1992}		& (3-8)$\times10^{8}$ \citep{wilman2005, 3c84_mass2013} &  $10^{45}$    \\
IC 310	   &FR1 & 0.0189	(80) \citep{bernardi2002}   	& (1-7)$\times10^{8}$ \citep{bettoni2003, ic310_science} &  $10^{44}$    \\
3C 264	   &FR1 & 0.0217	(95) \citep{smith2000}   	& 2.6$\times10^{8}$  \citep{van2016} &  $6\times10^{43}$    \\
PKS 0625-35&FR1/BL~Lac	& 0.05488	(220) \citep{jones2009}		& $3\times10^{9}$    \citep{bettoni2003} &  $5\times$$10^{41}$    \\
\bottomrule
\end{tabular}
\label{tab1}
\end{table}

\section{Observations of the TeV emitting radio galaxies} 

Among the more than 30 radio galaxies observed by the {\it Fermi}/Large Area Telescope (LAT) reported in the 
LAT 8-year Point Source List (FL8Y : https://fermi.gsfc.nasa.gov/ssc/data/access/lat/fl8y/), 
six have also been detected   at TeV energies (see Table \ref{tab1}). Most of these 
sources are  relatively nearby; the closest is Centaurus A at a distance of 3.7 mega-parsec, while 
PKS 0625-35\footnote{The source class for PKS 0625-35 is still under debate. 
The large-scale morphology  of the source looks like that of radio galaxies, while
 its optical spectrum indicates a BL Lac classification \citep{wills2004}.} at 220 mega-parsec 
 distance is the farthest TeV-detected radio galaxy. Except 
from M87 and PKS 0625-35, with a black hole mass of a few $10^{9}$ M$_{\odot}$, the 
$TeVRad$ have black hole masses of the order of a few  $10^{8}$ M$_{\odot}$, which is  smaller than  the 
average black hole mass of beamed radio galaxies a.k.a.\ blazars \citep{falomo2003}.  In fact, the closet 
$TeVRad$, Centaurus A,  hosts an even smaller black hole, (M$_{BH}$ $\sim$5$\times10^{7}$ M$_{\odot}$). 
The radio morphology of these sources resemble  blazars when viewed at larger angles. 
The beaming effects are fairly moderate  in $TeVRad$ \citep[apparent speed less than a few times the 
speed of light][]{lister2016,  mertens2016, meyer2015}.

The detection of the nearest $TeVRad$, Centaurus A core, at TeV energies 
was first reported  by H.E.S.S.
using more than 100 hours of observations taken during 2004-2006 \citep{aharonian2009}.  
The source 
has been later detected also at GeV energies by the  {\it Fermi}/LAT \citep{abdo2009}. 
{\it Fermi}/LAT  also discovered the spatial extension in the source, detecting $\gamma$-rays 
from both the core and from the extended giant lobes \citep{abdo2009}. The combined 
GeV-TeV spectrum of the source shows a clear excess at TeV energies and cannot be fitted 
via a single zone emission model \citep{sahakyan2013}. Moreover, the origin of the 
extended $\gamma$-ray emission  still remains an open question, although several models were proposed, 
i.e.  inverse-Compton scattering of relic radiation from the cosmic microwave 
background  and the infrared-to-optical extragalactic background light \citep{abdo2009}.

At a distance of $\sim$16~Mpc, M87 was the first TeV detected radio 
galaxy \citep{aharonian2003}. M87 is the brightest galaxy in the Virgo cluster. The source has been detected on several 
occasions  both in flaring and quiescent states \citep{veritas2009, magic2012, aharonian2006}. The fastest 
variability timescale observed in the source is $\sim$1~day \citep{aharonian2006}. Long-term 
variability on a few week timescales has also been reported for the source especially 
during quiescent states \citep{beilicke2012, magic2012}.

The {\it Fermi} telescope has detected 3C~84 since the beginning of its operation \citep{fermi_3c84}. 
{\it Fermi} observations suggest that the GeV photon flux of the source has been continuously rising 
since then (see Section \ref{tev_var} for details). 
In 2009, the source was also detected at TeV energies  \citep{aleksic2012, benbow2015}. In the end 
of 2016 and early 2017, the source has been through a dramatic flaring activity. During this epoch, 
the MAGIC telescopes detected the brightest flare ever observed in the source 
\citep[$L_{\gamma-ray}$ $\sim 10^{45}~erg~s^{-1}$,][]{magic2018} with a flux doubling timescale 
of $\sim$10~hours.

The radio galaxy IC~310 was first detected at TeV energies in 2009-2010 by the MAGIC telescope 
\citep{aleksic2010}. Variations on days to months timescales were observed in the source 
during this period. In November 12-13, 2012, a dramatic flare was observed in the source. 
During this exceptional variability phase, the observed flux doubling timescale of 
the observed TeV flares was as short as 5~minutes. This is the fastest variability 
timescale observed in a radio galaxy up to  now \citep{ic310_science}. This extreme flare is 
quite challenging to be explained within the framework of any existing theoretical  
mechanism (see Section \ref{emission_mech} for details). 

At a distance of 95 Mpc, 3C 264 is rather a new member in the family. 
Nearly 12 hours of VERITAS observations of the source between February 09, 2018 and  
March 16, 2018 confirms its detection at $\sim$5$\sigma$ level \citep{reshmi2018}.
TeV emission above 250~GeV from PKS 0625-35 was first reported in 2012 by H.E.S.S. \citep{dryda2015}. 
{\it Fermi} observations of the source report variations on monthly timescales \citep{fikazawa2015}. 
However, no variation has been seen seen at TeV energies. The source belongs to the category of low excitation 
line radio-loud AGN \citep{liner2006}. Radio observations find superluminal 
motion in the jet with apparent speeds up to $\sim$3~c  \citep{mueller2012}. The evidence of a  
one-sided jet and superluminal motion suggests the source to be more like a blazar.

\begin{figure}[t]
\centering
\includegraphics[scale=0.7, trim = 1 120 1 1, clip, angle=0]{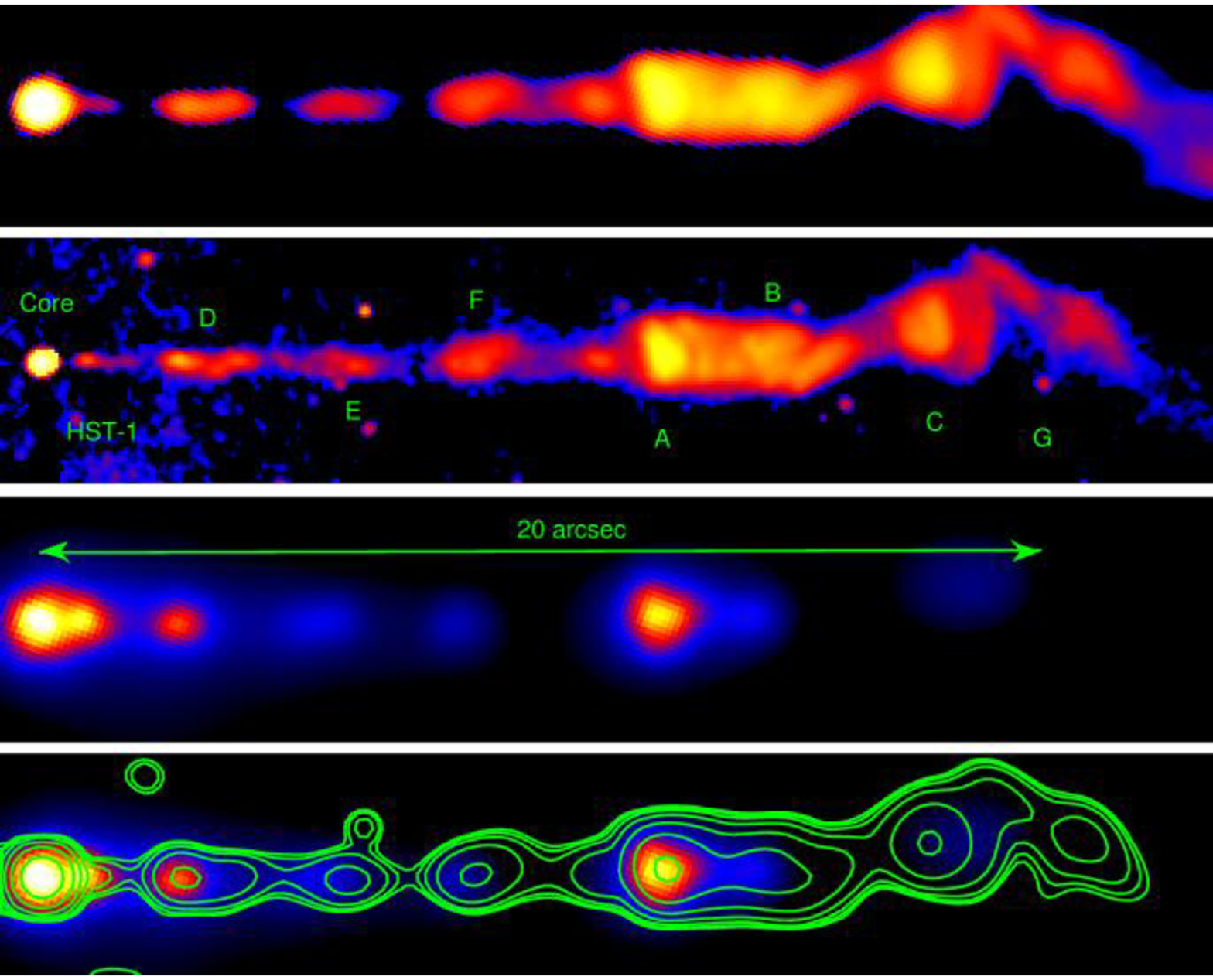}
\caption{Multi-wavelength view of the M87 jet. From top to bottom:  radio jet imaged by  Very Large 
Array (VLA), optical jet seen by the Hubble Space Telescope, and X-ray image taken by the Chandra 
X-ray Observatory. The optical image 
(middle) marks the position of various bright knots/stationary features observed in the jet.  
(Credit: \citet{marshall2002}, reproduced with permission \textcopyright AAS).}
\label{fig_m87jet}
\end{figure}

\section{Radio properties} 

\subsection{Parent population} 
All $TeVRad$ are Fanaroff and Riley type I 
(FR1) sources. Moreover, most of the TeV-detected blazars 
(high-frequency peaked BL Lacs, HBL, in TeVcat: http://tevcat.uchicago.edu/)  belong to the parent 
population of FR1 sources.  The jetted AGN (also known as radio-loud AGN) 
are classified as FR1 and FR2 \citep{fanaroff1974}  sources on the basis of 
their jet luminosity at 178~MHz radio frequency. 
It was found that the sources with luminosity $\geq$5$\times$10$^{24}~W Hz^{-1} Sr^{-1}$ 
belong to the FR2 class, while the rest were classified as FR1 type \citep{fanaroff1974}. 
The key difference between the two populations is their radio morphology.  
FR2s are more powerful radio galaxies, and their radio morphology   
is characterized by powerful edge-brightened 
double lobes with prominent hot spots,  while the FR1s are lower-power sources, which 
exhibit rather diffuse edge-darkened lobes. Moreover, FR2s tend to be found in less dense 
environments compared to FR1s. When seen at small inclinations, FR1 sources are BL~Lacs 
while FR2s are radio-loud quasars (an excellent  discussion about AGN unification can be 
found in \citet{urry1995}). 

It was found that at a given radio power, FR1s have fainter nuclear X-ray and 
ultra-violet (UV) emission \citep{hardcastle2006, baum1995} compared to FR2s, which indicates 
that FR1s might have fainter accretion disks.  For instance, the accretion 
rate for the 
brightest $TeVRad$ 3C~84 ($L_{VHE} = 10^{45} erg s^{-1}$) is well below 10$\%$ 
\citep[Eddington ratio $\sim$3$\times 10^{-4}$,][]{sikora2007}. 
The key question to answer is  if this 
corresponds to  different accretion modes in the two classes of sources \citep{fabian1995}. 
If yes, it would be interesting to assess how  it correlates to their observed TeV luminosity. 
Moreover FR1s may occupy less dense environments, enabling lower $\gamma\gamma$-pair production 
opacity for the TeV photons. 
In dense environments, high-energy photons might get absorbed via a process called 
pair production when they interact with lower-energy (optical/UV) photons by producing 
electron-positron pairs.

\begin{figure}[t]
\centering
\includegraphics[scale=0.55, trim = 1 1 1 1, clip, angle=90]{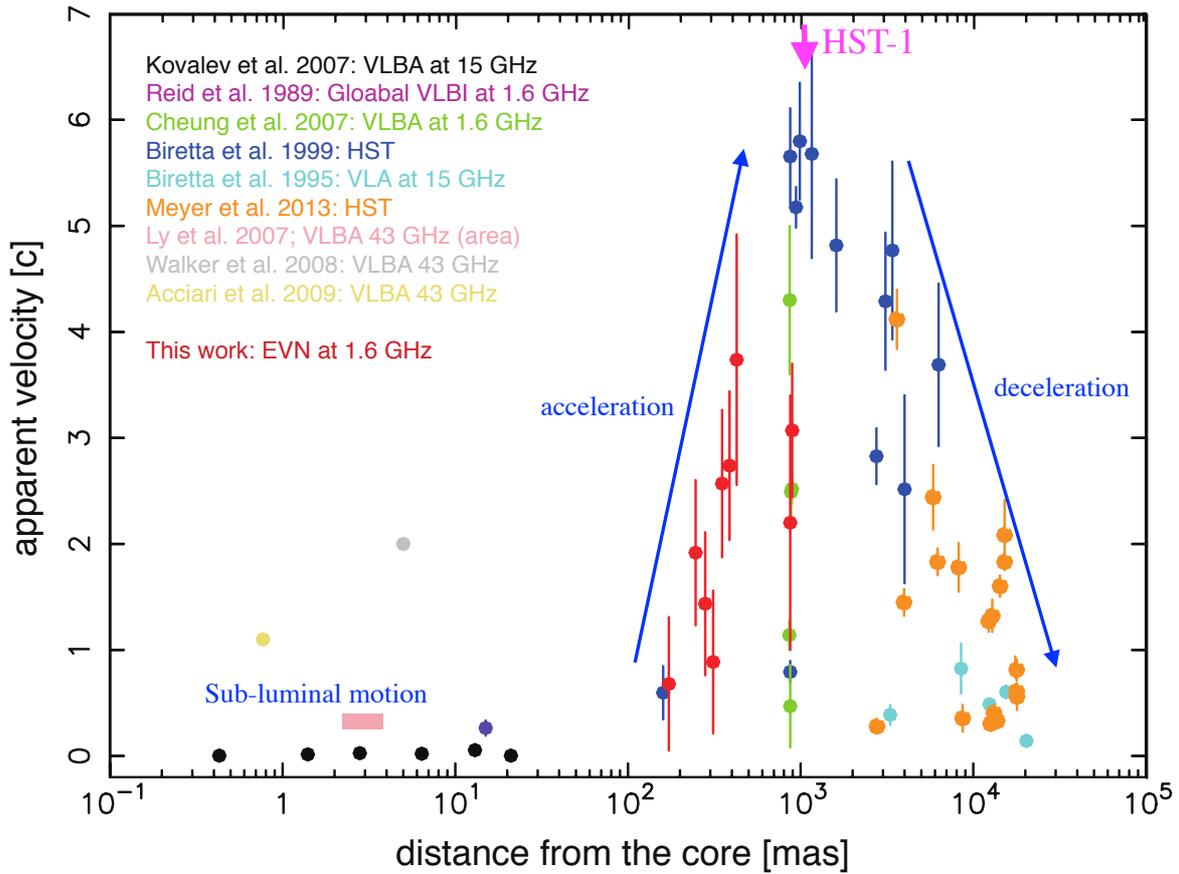}
\caption{Velocity field measurements in the  jet of M87 from parsec to kilo-parsec scales. The stationary feature HST-1 
is at a distance of $\sim$ 100~parsecs from the central engine. The highest speeds are measured close to the position of HST-1. 
(Credit: \citet{asada2014}, reproduced with permission \textcopyright AAS).}
\label{fig1b}
\end{figure}

\subsection{Jet kinematics - speeds on parsec to kilo-parsec scales}

The $TeVRad$ M87 is the best studied radio galaxy because of its 
extremely bright and spectacular jet extending 
up to several hundreds of kilo-parsecs. Figure \ref{fig_m87jet} depicts the kilo-parsec 
scale radio, optical, 
and X-ray images (from top to bottom) of the M87 jet. 
The high-resolution VLBI observations 
show very sub-luminal speeds ($\sim$0.01~c) on tens of parsec scales \citep{kovalev2007, mertens2016, asada2014}. 
Later there is a sudden increase in the apparent speeds at a distance of $\sim$100~parsecs just prior to the stationary feature HST-1. 
Superluminal components with speeds up to 6c are detected after HST-1 
(see Fig.\ \ref{fig_m87jet}), which is followed by a  smooth deceleration 
\citep{mertens2016, biretta1995}. A similar velocity  pattern has observed in 3C~264. The sub-luminal apparent speeds ($\leq 0.5~c$) 
seemingly receive  a kick at a distance of $\sim$100~parsec where  the jet has a stationary knot/feature. On kilo-parsec 
scales apparent speeds up to 7~c have been observed in the source \citep{meyer2015}.    
How exactly these velocity fields develop and evolve along the jet is still an open 
issue. The stationary features are thought to be  the sites of  recollimation shocks, which might have an 
interesting role in accelerating  the sub-luminal  motion to super-luminal speeds.

There has been no dramatic change noticed in the apparent speed of the moving 
features in the other $TeVRad$ as a function 
of distance from the central engine.  For instance, 
radio VLBI and X-ray observations suggest sub-luminal motion (speeds up to 0.5~c) both in 
the parsec and kilo-parsec 
scale jet of Centaurus A \citep{mueller2014, tingay2001,hardcastle2003}. Similarly the observed speeds on parsec 
and kilo-parsec scales are sub-luminal ($\sim$0.2 - 0.9~c) in 3C 84 \citep{hodgson2018, lister2013, dhawan1998, jorstad2017}, 
suggesting a viewing angle $\geq$45$^{\circ}$. 
The one-sided blazar like jet of IC~310 constrains the   a viewing angle 
to be $\leq$38$^{\circ}$ implying a moderate beaming \citep{kadler2014}.  VLBI observations 
 measure an apparent speed of $\sim$0.9--3~c in  the PKS 0625-35 jet, which constrains the viewing angle in the source  
 to be $\leq$13$^{\circ}$ (Angioni et al.\ 2018, submitted).

\begin{figure}[t]
\centering
\includegraphics[scale=0.45, trim = 1 1 1 1, clip]{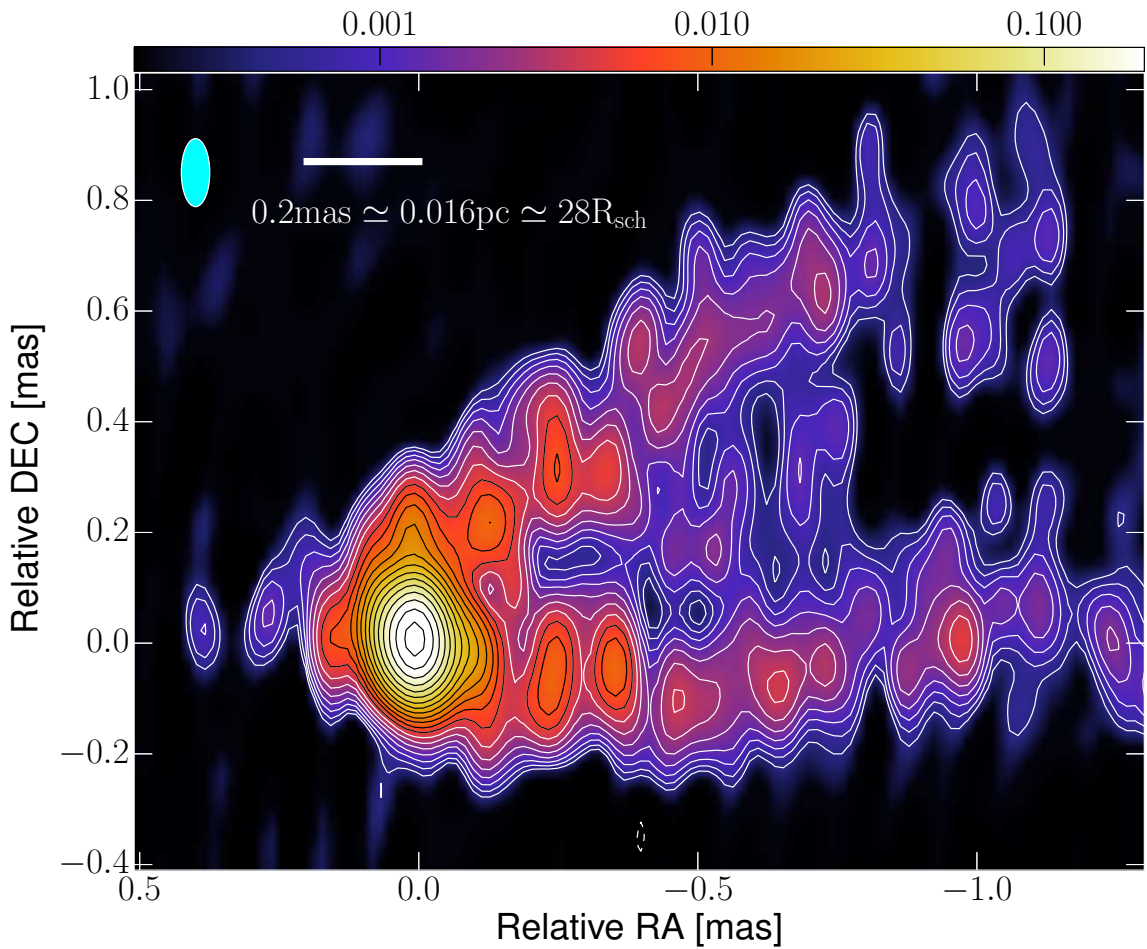}
\caption{Stacked high-resolution radio image of M87 at 86 GHz frequency 
(Credit: \citet{kim2018}, reproduced with permission \textcopyright ESO). The high-resolution radio imaging 
technique allows  a zoom up to a scale of $\leq$10 gravitational radii in the source frame.  }
\label{fig4}
\end{figure}

Structured and stratified velocity patterns have also been observed in 
the parsec-scale jets of some of the $TeVRad$. For instance, a study by \citet{mertens2016} 
report  a stratification in the M87 jet  with a bright transverse velocity
component. The jet flow in the source has a slow, mildly relativistic layer, sheath,  
(apparent velocity $\sim$0.5c) and a fast spine (apparent velocity $\sim$0.9c).
The study also discovered  a significant difference in the apparent speeds observed 
in  the northern (0.5~c) and southern (0.2~c) limbs of the jet (the bright limbs 
in the source can be seen in Fig.\ \ref{fig4}).  Likewise, limb brightening was  
discovered in the 3C 84 jet using the  space-VLBI observations \citep{3c84_radioastron}. 
Hints of differential motion were also observed in Centaurus A \citep{mueller2014}.

\subsection{Zooming into the event horizon scales} 
AGN jets are the most spectacular structures in the Universe and 
represent the ``extreme" and ``largest" members of a wide 
family of jet-powered phenomena, including $\gamma$-ray bursts (GRBs), X-ray binaries (XRBs), 
and young stellar objects (YSOs). 
Over the past few decades, there has been remarkable progress in our 
understanding of AGN jet physics especially because of the development of 
high-resolution VLBI (for more details refer to the recent review on 
millimeter-VLBI by \citet{boccardi2017}. Millimeter-VLBI observations offer an angular 
resolution of a few tens of micro-arcseconds, which allows a probe of the inner jet region 
on scales $\leq$50 gravitational radii ($R_g$).

Given its proximity and fairly bright  jet at millimeter radio frequencies,  M87 is the best studied 
source using high-resolution VLBI \citep{hada2011, kim2018, doelman2012, asada2014, mertens2016}. 
Global millimeter VLBI observations at 86~GHz radio frequency resolve a limb-brightened 
outflow down to $\sim$15~$R_g$ \citep{kim2018}. A high-resolution image of M87 
taken at 86~GHz radio frequency is illustrated in Fig.\ \ref{fig4}. 
The Event Horizon Telescope (EHT) observations at 230~GHz and 345~GHz frequencies 
constrain the size of 
the event horizon of the central black hole to be less than 10~$R_g$ 
\citep{doelman2012, lu2014}. These high-resolution images provided a unique 
tool to test the different jet-launching models \citep{doelman2012, hada2011}

Another interesting study is made using the space-VLBI observations. 
The space-VLBI observations of 3C~84 at 22~GHz radio frequency  using the {\it RadioAstron} mission 
probed the inner jet profile up to 100~R$_g$ \citep{3c84_radioastron}. The study 
discovered a very broad jet base with a transverse radius larger than about 250~$R_g$ 
implying the jet  to be more likely launched from the accretion disk. More interestingly, 
GMVA observations at 86~GHz radio frequency discovered  a double-nuclear structure at the 
jet base (Oh et al. 2018, in preparation).

High-resolution VLBI observations therefore offer an unparalleled opportunity to probe and 
pinpoint the high-energy emitting sites in the immediate vicinity of the central black hole. 
Moreover, high-resolution polarization imaging is a unique technique to unravel 
the magnetic field topology on scales less than a few hundreds of gravitational radii. 
The tiny magnetized regions in the extended jets are the potential 
sites of high-energy emission \citep{rani2018, rani2015, marscher2008}.

\begin{figure}
\centering
\includegraphics[scale=0.6, trim = 1 1 1 1, clip]{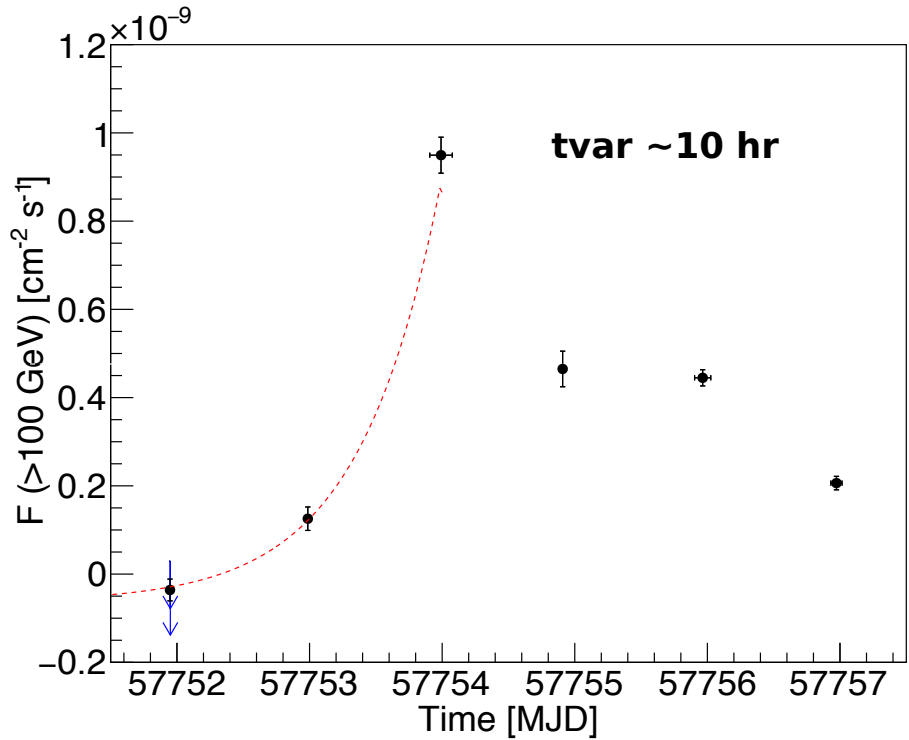}
\caption{TeV light curve of 3C~84 observed by the MAGIC telescope \citep{magic2018}. The 
fastest flare observed around 01-01-2017 has a flux doubling timescale of $\sim$10~hr.  
Long-term variations at $\gamma$-rays are shown in Fig.\ \ref{fig2}. }
\label{fig3}
\end{figure}

\section{TeV variability}
\label{tev_var}
\begin{figure}[t]
\centering
\includegraphics[scale=0.7, trim = 1 1 1 1, clip]{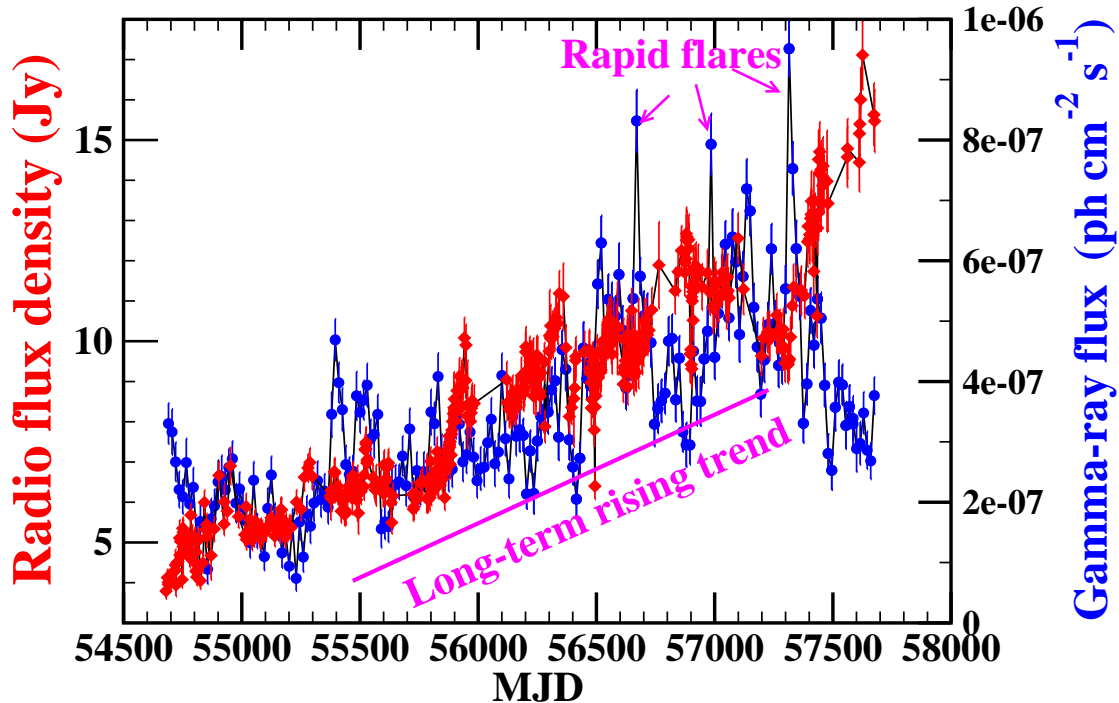}
\caption{Gamma-ray photon flux (blue) and radio 230~GHz (red) light curve of the 
$TeVRad$ 3C~84 at E$=$0.1-300~GeV. The $\gamma$-ray light curve is produced by 
adaptive binning analysis method following \citet{lott2012} 
using the {\it Fermi}/LAT data from August 2008 until September 2017 . The radio 230~GHz light curve is from sub-millimeter array (SMA) 
AGN monitoring program (details are refereed to \citet{hodgson2018}). 
Rapid flares superimposed on top of the long-term rising trend can be seen both at 
$\gamma$-ray and radio frequencies. }
\label{fig2}
\end{figure}

Like blazars, $TeVRad$ exhibit variability on multiple timescales i.e. sub-hour to days timescale 
flares are accompanied by long-term outbursts. For instance, multiple modes of flaring 
activity can be seen in the photon flux light of 3C~84 observed by the {\it Fermi}/LAT. 
Figure \ref{fig2} presents the rapid flares superimposed on top of the long-term rising 
trend in the {\it Fermi} photon flux light curve (in blue).  In the end of 2016, several extremely 
bright and rapid flares were observed by {\it Fermi}/LAT. The fastest flares seen by {\it Fermi}/LAT 
have flux doubling timescale of 9-12~hrs.
In 2016-2017, the source was also detected multiple times 
at TeV energies by the MAGIC telescope \citep{magic2018}. During the course of this period, 
MAGIC detected the fastest and most luminous flare from 3C~84. An extremely bright  
($L_{\gamma}\simeq 10^{45}~erg~s^{-1}$) flare with a flux doubling 
timescale of $\sim$10~hours was observed on January 01, 2017 \citep{magic2018}. 
Flux variations on multiple timescales were also observed in IC~310 and M87.

\begin{figure}[h]
\centering
\includegraphics[scale=0.5, trim = 1 1 1 1, clip]{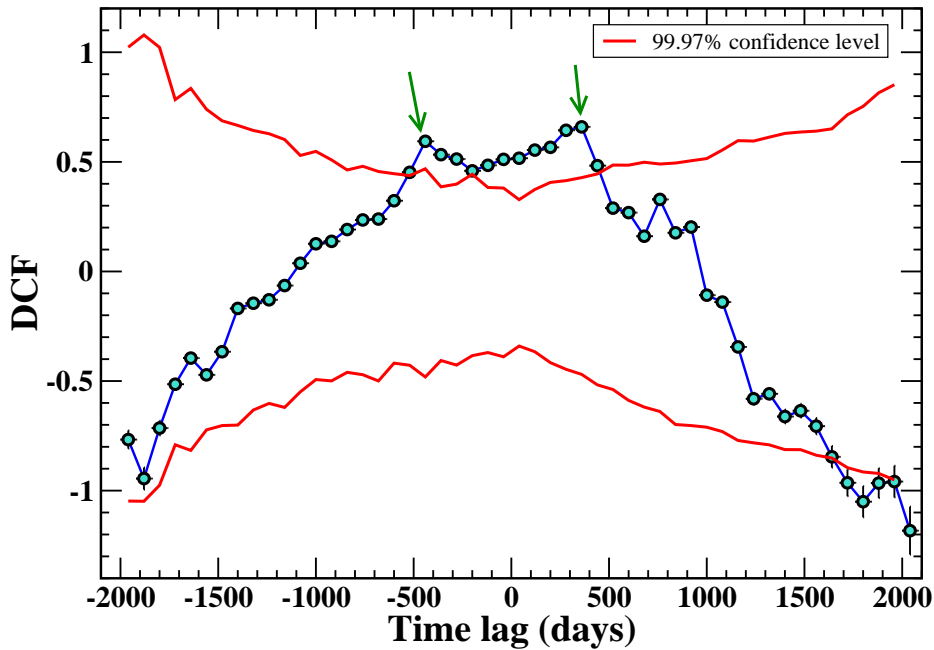}
\caption{Cross-correlation analysis curve of the $\gamma$-ray and radio flux variations 
observed in 3C~84 (see Fig. \ref{fig2}). The two peaks  at $-$500 and $+$400 days  (green arrows) imply  
$\gamma$-rays flux variations  leading and lagging those at radio. A detailed analysis 
suggests the presence of multiple $\gamma$-ray emitting sites in the source \citep{hodgson2018}.}
\label{fig_dcf}
\end{figure}

Extremely short variability timescales in $TeVRad$ have attracted the attention of  
astrophysicists, as the observed timescales can reach down to the light crossing time of 
the black hole event horizon, $t_{BH} = r_{g}/c$, where $r_g$ is the gravitational radius 
of the black hole and $c$ is the speed  of light. For instance, the observed variability timescale 
(t$_{var}$ = 5 min) in IC~310 is much smaller than the event horizon light crossing timescale, 
$t_{BH}$ = 25 min \citep{ic310_science}. The  fastest variability timescale 
observed in M87 (t$_{var}$ $\sim$ 1~day) is comparable with the event horizon 
light crossing timescale ($t_{BH} \sim$ 60-90 min). 
Variations on sub-horizon scales have also been 
observed in blazars \citep{albert2007}. However in case of blazars, one can always 
argue that beaming effects make the observed timescale  appear shorter. This requires 
the Doppler factor to be greater than 50, which has rarely been observed \citep{lister2016, jorstad2017}. 
It is also important to note that the timescale of the flux variations in blazars 
is also affected by time-dilation in the frame co-moving with the jet. Since the Doppler 
factor, $\delta$ $\simeq$ $\Gamma_{jet}$ (Lorentz factor in the jet), the projection 
effects (used to explain the ultra-fast variability in blazars) cancel out.

Understanding the physical processes responsible for the origin of 
the ultra-fast TeV flares is a key challenge in high-energy astrophysics. 
Sub-horizon scale variations and extremely luminous  flares, especially in 
the absence of strong beaming effects, require (i)  emission regions of extremely 
high electromagnetic energy, (ii) efficient particle acceleration mechanisms which could 
channel the electromagnetic energy to particle energy, 
and (iii) less-dense photon environments so that the emitted TeV photons do not 
get self-absorbed by $\gamma \gamma$-pair production.

\section{High-energy emitting sites} 
Locating the $\gamma$-ray emitting sites is a key challenge not only in case of  
radio galaxies but also for blazars. When the emission region is closer to the central 
black hole, there is higher proportion of electromagnetic energy available in the jet, making 
it easy to account for the observed extremely bright $\gamma$-ray flashes. However if the 
high-energy photons are produced too close to the black hole, the chances for them  to escape  through 
the rich photon environments are low. 
Nevertheless, at least some 
of these high-energy photons seem to survive even through dense environments. 
Photons produced further down in the jet have higher chances of not being absorbed; however 
accelerating  particles to extreme high-energy is a quite challenge there 
(see Section \ref{emission_mech} for more details).

The sub-horizon scale emitting regions, detected in IC~310, 3C~84, and M87,  
could either be very close to the black hole or further down in the jet. 
The TeV flare in IC~310 was interpreted as  originating  
from the magnetospheric gap of the black hole (see Section \ref{emission_mech}) as 
it is hard to achieve the high luminosity and rapid variability further down 
in the jet. The rapid TeV flare in 3C~84 was also proposed to be produced 
in the magnetospheric gap; however achieving the extremely high luminosity requires 
an extremely fast rotating black hole  and magnetic field higher than its equipartition 
value \citep{3c84_tev}. However these studies  lack a conclusive observational 
evidence for 
the emission to be produced close to the black hole.  
Multi-frequency observations  of M87 during an episode of extremely rapid flares provided 
an opportunity to study the TeV emitting regions in great detail.  Strong 
TeV flares in the source were accompanied by an increase of the  flux from the nucleus 
observed at 43~GHz radio frequency \citep{acciari2009}, suggesting   
the emission region to be located within the radio core ($\leq$50 gravitational radii).
{\it The extreme compactness of the TeV emission region and 
the observed TeV-radio connection in M87 conclusively infer that the VHE emission is indeed 
produced very close to the black hole \citep{acciari2009}}.

Hints of multiple $\gamma$-ray emitting regions have also been observed in 
$TeVRad$. For instance, a study performed using the KVN (Korean VLBI Network)  
and $\gamma$-ray observations indicate  the presence of multiple $\gamma$-ray 
emitting regions in 3C~84 \citep{hodgson2018}. Figure \ref{fig2} presents a 
comparison of the flux variations observed at $\gamma$-ray and 230~GHz radio 
frequencies. The long-term  rising trend is present in the flux variations  
seen both at radio and $\gamma$-ray frequencies. However the rapid flares do not have a one-to-one 
correspondence at the two frequencies. A formal cross-correlation analysis indicates 
two prominent peaks at around  $-$500 and $+$400~days, which implies that $\gamma$-rays 
variations lead those at radio by $-$500~days and also lag behind by $+$400 days 
(see Fig.\ \ref{fig_dcf}). A 
detailed investigation of the flux density variations in the spatially separated emission 
regions of the jet indicates the presence of multiple $\gamma$-ray emitting sites in the 
source. The  study  reported that there are two active regions  producing 
$\gamma$-rays in the source,  one very close to the black hole and another at a distance of $\sim$4 parsec 
in the source frame.

{\it Fermi} has detected $\gamma$-rays both from the core and the extended lobes 
of the  nearest $TeVRad$ Centaurus A \citep{abdo2009}. The source has a spatial 
extension of About 
9 degrees across the sky, which corresponds to $\sim$600 kilo-parsec, and  the $\gamma$-ray 
morphology  is quite similar to its  radio jet morphology, which implies $\gamma$-rays 
coming from both the central AGN core and extended kilo-parsec scale lobes. 
Given the detection of extended $\gamma$-ray emission at GeV energies and the absence 
of TeV variability, the extension of TeV emission cannot be discarded. It is 
important to note that {\it Fermi}/LAT has detected extended GeV emission also from 
Fornax~A. It might be interesting to see for how many of these sources  spatial extension 
can detected by using the future Cherenkov Telescope Array.

\section{Particle acceleration mechanisms} 
\label{emission_mech}

 The observed extremely rapid variability and enormously high $\gamma$-ray luminosity are 
clearly challenging to achieve in astrophysical jets as they require (i) particle acceleration 
to extreme high energies in presence of strong radiative loss (ii) variation on 
sub-horizon  timescales, and (iii) conversion of large volumes of electromagnetic 
energy  with enormous efficiency.  Several mechanisms have been proposed to 
address the TeV challenge in astrophysical sources. Some recent ideas are summarized 
below.

The most common mechanism is relativistic shock i.e.\ the flow is  accelerated to 
supersonic speeds and strong shocks  are formed in the relativistic outflow 
because of (i) faster moving outflow runs into slower flow resulting in 
the formation of internal shock, (ii) the outflow encounters an obstacle 
(perhaps a molecular cloud) on its way, (iii) any other perturbation happening 
at the base of the jet could also form a shock wave to form and propagate 
down the jet. Relativistic shocks \citep{marscher1985} seem to be effective 
in low-magnetization regimes and could explain the observed long-term variability from 
AGN fairly well \citep{rani2013, rani2015, rani2018, chidiac2016, hodgson2017, 
karamanavis2016, marscher2008}. 
However, the luminous ultra-fast TeV  flares observed from a number 
of AGN \citep{aharonian2007, albert2007, 3c84_tev, ic310_science} are difficult  be  explained 
via the relativistic shock model especially in the absence 
of substantial Doppler beaming,    and  
therefore challenge our understanding of particle acceleration processes.

\begin{figure}[t]
\centering
\includegraphics[scale=0.6, trim = 1 1 1 1, clip, angle=-90]{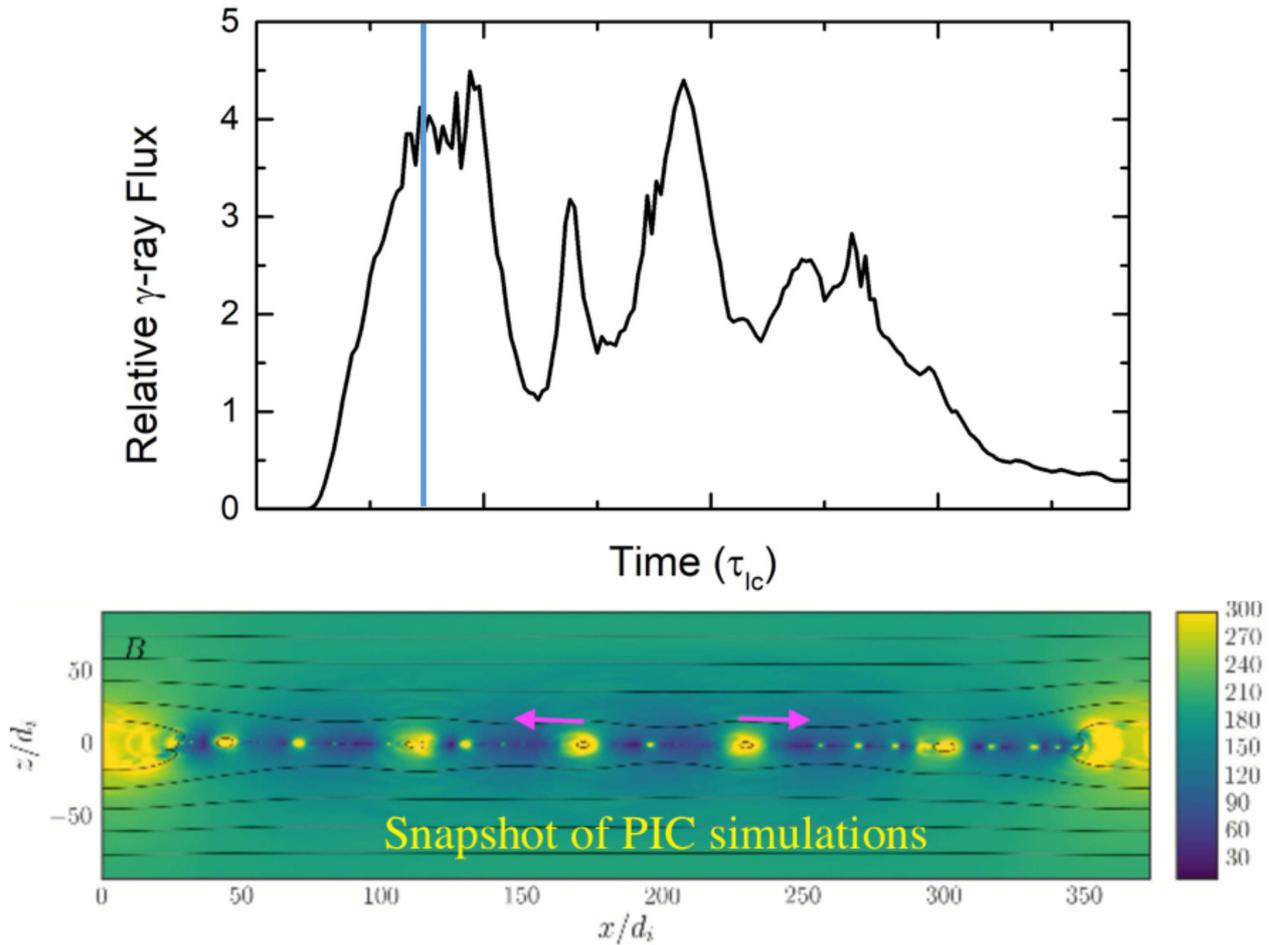}
\caption{Radiative signatures of relativistic magnetic reconnection
by an integrated modeling relying on first-principle particle-in-cell
(PIC) kinetic simulation (Credit: Haocheng Zhang \cite{zhang2018}). {\it Top panel:} $\gamma$-ray light curve in  units of 
light crossing time scale. {\it Bottom  panel:} reconnection layer in the unit of ion skin depth. }
\label{fig_reconnection}
\end{figure}

Magnetic reconnection, a process by which magnetic energy is transferred to particle heat, bulk energy 
and non-thermal energy, is proposed to be an effective mechanism 
for energy dissipation \citep{petropoulou2016, giannios2013, petropoulou2018, kagan2015, 
zhang2018} 
in magnetically dominated relativistic outflows. During  magnetic reconnection, opposite polarity 
magnetic field  lines exchange partners at ``X-points" leading to breaking of flux freezing and change in 
magnetic field topology, which results in an  explosive  release of energy. For instance, solar 
flares seem to be powered by magnetic reconnection. However this picture is far too complex 
in case of relativistic outflows. Nevertheless, there has been great advancement in simulations 
to understand the relativistic magnetic reconnection.   The particle-in-cell  (PIC) magnetic reconnection simulations 
suggest that ultra-relativistic plasmoids generated by magnetic 
reconnection \citep{petropoulou2016, giannios2013}  can power the ultra-fast 
flares observed from TeV AGN. An example $\gamma$-ray light curve (top) and snapshot of  relativistic magnetic 
reconnection (bottom)
by an integrated modeling relying on first-principle 
PIC kinetic simulation and an advanced polarized radiation transfer simulation from \citet{zhang2018}  
are illustrated in Fig. \ref{fig_reconnection}. 
The study reports  that plasmoid coalescences in the magnetic 
reconnection layer  can produce highly variable multi-wavelength light curves, and also can explain 
polarization degree and polarization angle variations.

A magnetospheric gap model was invoked to explain the sub-horizon scale flux variations in 
$TeVRad$ \citep{ic310_science, 3c84_tev, hirotani2016, rieger2011}. It has been proposed 
that a  rotating black hole embedded in a strong magnetic filed 
(of the order of 10$^{4}$ - 10$^{5}$~G) will induce an electric field, $E$, corresponding to a 
voltage drop across its event horizon, which facilitates an efficient  electromagnetic extraction 
of the rotational energy of the black hole (for more details, see the review by 
\citet{rieger2011}), 
and could be an efficient mechanism to power the TeV flares in AGN. The maximum extractable 
energy ($L_{magnetosphere}$) from the black hole however depends on the details of the gap 
configuration. For instance, $L_{magnetosphere}$ scales as the height of the magnetospheric 
gap ($h$) i.e.\ i.e.\ $L_{magnetosphere}$ $\sim L_{BZ}(h/r_g)^{\beta}$, where 
$L_{BZ}$ is the maximum  power from a fast rotating black hole \citep{BZ1977} and $\beta$ 
depends on the gap setup.  
The size of the gap can be constrained by the observed TeV variations 
($c.\delta t$ $\leq$ $r_g$), which further reduces the maximum $L_{magnetosphere}$ 
\citep[details are refereed to][]{katsoulakos2018, rieger2011}. Another important 
aspect of this model is the absorption of the emitted TeV photons by the 
thermal photon environment of the black hole. In order for the TeV photons to survive 
through the accretion disk environment, the magnetospheric gap model requires an 
under-luminous or radiatively inefficient accretion flow to avoid  $\gamma\gamma$-pair 
production. This enforces accretion rates to be below 1$\%$ of the Eddington scale. Putting this in the 
context of current observations, the model could explain the observed TeV flares in M87 
fairly well; however the minute-scale TeV flare in IC~310 seems less likely 
to have a magnetospheric gap origin. The observed TeV luminosity in 3C~84 is also hard to 
match in the framework of the magnetospheric gap model unless the magnetic field threading the 
accretion disk is a factor of 1000 greater than its equipartition value.

{\it Magnetoluminescence} \citep{blandford2017, blandford2015} has been recently proposed 
to explain the observed TeV flares from astrophysical sources. The basic idea is that 
rotating at the center of an AGN leads to writhing of  magnetic flux ropes resulting in 
tangled and knotted  magnetic field lines. The tightly wound magnetic flux ropes especially 
in the inner jet are highly dynamically unstable. The tangled/knotted flux lines when opened up 
or untangled could produce rapid $\gamma$-ray flares. It has been suggested that this phenomenon 
includes an implosion rather than explosion \citep{blandford2015}. As a consequence, particles 
could be accelerated to significantly higher energies as the dissipating  and cooling volume 
could be re-energized by the medium external to it.  
Electromagnetic detonation \citep{blandford2015} is proposed as an alternative possibility, 
which might 
accelerate particles at the transition of high-magnetized outflow  into 
low-magnetized outflow and energy-flux. Particles  are subjected to unbalanced 
electric field at the transition and get accelerated (positive and negative charged 
particles are accelerated in opposite direction) to produced $\gamma$-rays. 
In comparison to magnetic reconnection which require the magnetic flux to channel into 
a small volume,  electromagnetic detonation will have a larger volume for the 
conversion of electromagnetic energy into particle energy.

\section{Future directions} 
Pinpointing the sub-horizon scale $\gamma$-ray emitting sites in kilo-parsec to mega-parsec scales 
AGN jets is a key challenge in high-energy astrophysics. A 
coordinated multi-wavelength and multi-messenger effort is needed to better understand it. 
The TeV emitting radio galaxies ($TeVRad$) are a particularly interesting class of AGN to unravel
the mystery of the origin of $\gamma$-ray emission.  Some important considerations are: 
\begin{itemize} 
\item $TeVRad$   are nearby AGN.  The  farthest detected $TeVRad$ is at a 
distance of 220~Mpc. Using the millimeter-VLBI observations, one can probe regions in the immediate 
vicinity of the central black hole i.e.\ down to a scale of 
less than 50 gravitational radii in the source frame (for more details about the imaging 
capabilities of the current and near-future high-resolution VLBI observations, check 
\citet{boccardi2017}). The mm-VLBI observations will provide an ultimate opportunity to 
probe the compact regions closer to the black hole, which are  the potential sites of 
high-energy emission. High-resolution polarization imaging will probe the magnetic field strength 
and configuration of these compact emission regions, which is an essential piece of information 
in order to understand the high-energy acceleration processes. 

\item Being seen off-axis, $TeVRad$ offer a unique opportunity to transversely resolve  the 
jet. Using  high-resolution VLBI one can transversely 
resolve the fine scale jet structure and can probe their flux intensity and polarization 
variations. VLBI astrometry observations could provide the exact location of the core. Having 
that, one could combine the high-resolution VLBI imaging and multi-wavelength observations to 
exactly pinpoint the location of $\gamma$-ray emitting sites.

\item There have been great advances in the theory and simulation front. Several details 
about plasma physics under extreme conditions can be  better understood via simulations i.e.\ 
radiative signatures of relativistic magnetic reconnection \cite{petropoulou2016, giannios2013, 
petropoulou2018, kagan2015, zhang2018}, formation of jets and expulsion  of magnetic 
flux from the central black hole \cite{mckinney2013, mad2014, liska2018}, high-energy 
polarization signatures of 
shocks and reconnection \cite{zhang2018, zhang2017, marscher2014}. A comparison of 
predictions from simulations with measurements will be crucial to understand 
the physical processes happening around black holes.

\item Cherenkov Telescope Array \citep{cta2017} with its sensitivity better than the existing GeV/TeV 
instruments and spatial resolution of the order of a couple of arc-minutes at TeV energies will 
provide  observations up to 300~TeV. CTA will 
detect many more misaligned AGN and will also  help us in separating nuclear and off-nuclear
components especially for nearby sources. Having the potential to probe deeper 
into the spectral and variability characteristics of AGN, CTA will eventually revolutionize
our understanding of the TeV sky.

\item The question of what powers $\gamma$-ray AGN flares is ultimately related to the 
energy dissipation mechanism at work. High-energy polarization observations will deliver 
an ultimate test for probing the energetic particle-acceleration processes and emission 
mechanisms: (i) leptonic versus hadronic models and (ii) shocks versus magnetic reconnection.
High-energy polarimetry missions  are on their way to unravel how the most efficient 
particle accelerators in the Universe work. For instance, the  All-sky Medium Energy 
Gamma-ray Observatory (AMEGO: https://asd.gsfc.nasa.gov/amego/) will offer MeV polarization 
observations of the $\gamma$-ray bright AGN. The Imaging X-ray Polarimetry Explorer 
(IXPE: https://wwwastro.msfc.nasa.gov/ixpe/index.html) will observe polarization 
signatures from X-ray bright AGN.

\item The recent observations of  coincidence of a high-energy  neutrino event 
 with a flaring AGN, TXS 0506+056,  indicates AGN as  potential sources 
of high-energy neutrinos \citep{icecube2018}. Multi-wavelength and multi-messenger 
observations of similar events will test if AGN jets are powerful sources of 
extra-galactic neutrinos and put constraints on lepto-hadronic models. 

\end{itemize}

\bigskip 

\noindent 
{\bf Given the flood of high quality data from the current and  upcoming missions and 
great advances in theory/simulation, it is an exciting time for astronomers. There are 
several major discoveries to be made.}

\bigskip

\acknowledgments{BR acknowledges the help of Dave Thompson, Chris Shrader, Lorentzo Sironi, 
Benoit Lott, Zorawar Wadiasingh, and Alice Harding for  fruitful discussions 
and their comments that improved the manuscript. 
This research was supported by an appointment to the NASA Postdoctoral Program
at the Goddard Space Flight Center, administered by Universities Space Research Association 
through a contract with NASA.}
 

\end{document}